\documentclass{article}

\usepackage{arxiv}

\usepackage[utf8]{inputenc} 
\usepackage[T1]{fontenc}    
\usepackage{hyperref}       
\usepackage{url}            
\usepackage{booktabs}       
\usepackage{amsfonts}       
\usepackage{nicefrac}       
\usepackage{microtype}      
\usepackage{graphicx}
\usepackage{natbib}
\usepackage{doi}
\usepackage{appendix}
\usepackage{scrextend}
\usepackage[labelfont=bf]{caption}
\usepackage[ruled]{algorithm2e}
\usepackage{etoolbox}
\usepackage{svg}
\usepackage{setspace}
\usepackage{float}
\usepackage{subfloat}
\AtBeginEnvironment{algorithm}{\setstretch{1}}
\usepackage{statcommands}

\DeclareMathOperator*{\argmin}{arg\,min}
\newcommand{\cl}{{\,:\,}}
\allowdisplaybreaks

\title{A Comparison of Bayesian Inference Techniques for Sparse Factor Analysis}

\author{ Yong See Foo \\
    School of Mathematics and Statistics\\
	University of Melbourne \\
	Parkville, VIC, 3010, Australia \\
	\texttt{yongsee.foo@unimelb.edu.au} \\
	\And
	Heejung Shim \\
	School of Mathematics and Statistics and\\ 
	Melbourne Integrative Genomics\\
	University of Melbourne \\
	Parkville, VIC, 3010, Australia \\
	\texttt{heejung.shim@unimelb.edu.au}
}

\renewcommand{\headeright}{}
\renewcommand{\undertitle}{}

\begin{document}
\maketitle

\begin{abstract}
Dimension reduction algorithms aim to discover latent variables which describe underlying structures in high-dimensional data. Methods such as factor analysis and principal component analysis have the downside of not offering much interpretability of its inferred latent variables. Sparse factor analysis addresses this issue by imposing sparsity on its factor loadings, allowing each latent variable to be related to only a subset of features, thus increasing interpretability. Sparse factor analysis has been used in a wide range of areas including genomics, signal processing, and economics. We compare two Bayesian inference techniques for sparse factor analysis, namely Markov chain Monte Carlo (\textsc{mcmc}), and variational inference (\textsc{vi}). \textsc{vi} is computationally faster than \textsc{mcmc}, at the cost of a loss in accuracy. We derive \textsc{mcmc} and \textsc{vi} algorithms and perform a comparison using both simulated and biological data, demonstrating that the higher computational efficiency of \textsc{vi} is desirable over the small gain in accuracy when using \textsc{mcmc}. Our implementation of \textsc{mcmc} and \textsc{vi} algorithms for sparse factor analysis is available at \url{https://github.com/ysfoo/sparsefactor}.
\end{abstract}

\section{Introduction}
Dimension reduction techniques have been widely used for inferring and explaining an underlying structure in high dimensional data. One of these techniques is \emph{factor analysis}, which linearly maps high dimensional data onto a lower dimensional subspace. This is achieved by finding a set of latent variables, known as \emph{factors}, such that the observed variables may be represented by linear combinations of these factors. The aim of dimension reduction is realised by using a number of factors much smaller than the number of observed variables.

In some applications, it is desirable for each factor to be associated with only a subset of the observed variables. In other words, the factor loadings, which quantify the weighting of each variable on each factor, are expected to be sparse. \emph{Sparse factor analysis} is an extension of factor analysis that allows such sparsity to be captured. The benefit of sparse factor analysis is its increased interpretability of the inferred factors, as each factor is encouraged to have only a few significant loadings.

Sparse factor analysis has been applied to the analysis of gene expression data~\citep{West2003, sabatti2006, fatf}. One of the aims of such analyses is to infer gene regulatory networks, i.e. to identify sets of genes each regulated by a shared biological pathway. Thus, the use of sparse factor models is appropriate, as it allows the interpretation of factors as biological pathways, which each regulate a small number of the genes. Recent extensions of sparse factor analysis in genomics include~\citet{gao2016, hore2016, fsclvm, mofa, ebmf}.

Bayesian approaches have modelled the sparsity of factor loadings by using sparsity-inducing priors such as a ``spike and slab prior'' \cite{West2003}.
Markov chain Monte Carlo (\textsc{mcmc}), which relies on sampling from the posterior distribution, has been typically employed for Bayesian inference in sparse factor analysis. On the other hand, the recent extensions of sparse factor models in \citet{hore2016, fsclvm, mofa, ebmf} have used variational inference (\textsc{vi}). \textsc{vi} reformulates the inference problem to an optimisation problem of finding an approximate distribution that resembles the posterior distribution. It is known that \textsc{vi} tends to be faster than \textsc{mcmc}, but it does not provide theoretical guarantees of finding the exact posterior distribution, which \textsc{mcmc} provides \citep{viblei}.

We aim to investigate the relative strengths and weaknesses of \textsc{mcmc} and \textsc{vi}, when applied to sparse factor models with a spike and slab prior. We derive and implement \textsc{mcmc} and \textsc{vi} algorithms and assess a trade-off between accuracy and computational efficiency using both simulated and biological data. \citet{manu} performed a similar comparison, but they used a relaxed sparsity prior for their \textsc{vi} algorithm, instead of the exact spike and slab prior. Our work differs from \citet{manu} as we consider a slightly more flexible sparse factor model, and derive a \textsc{vi} algorithm for the exact spike and slab prior. Our comparison results show that the higher computational efficiency of \textsc{vi} is desirable over the small gain in accuracy when using \textsc{mcmc}, provided that sufficient \textsc{vi} trials are run. 
Our implementation of the \textsc{mcmc} and \textsc{vi} algorithms for sparse factor models is available at \url{https:github.com/ysfoo/sparsefactor}.

\section{The sparse factor model}
Given $N$ observations $\mathbf{Y} = [\boldsymbol{y}_1, \boldsymbol{y}_2, \ldots, \boldsymbol{y}_N]$ each with $G$ features, the sparse factor model describes the data using $K$ factors with a loading matrix $\displaystyle \mathbf{L}\in \mathbb{R}^{G\times K}$ and activation matrix $\mathbf{F}\in \mathbb{R}^{K\times N}$ such that 
$\mathbf{Y} = \mathbf{L}\mathbf{F} + \mathbf{E}$, where $\mathbf{E}\in \mathbb{R}^{G\times N}$ is a matrix of random errors. In the context of gene expression, $\mathbf{Y}$ represents gene expression data across $N$ samples, each measured on $G$ genes. A possible interpretation of the $K$ factors is to view them as biological pathways which regulate gene expression. By assuming independent normal errors with feature-specific variance, the distribution of $\mathbf{Y}$ is given by
\begin{equation}
p\giventhat*{\boldsymbol{y}_{\cdot j}}{\mathbf{L}, \mathbf{F}, \boldsymbol{\tau}} = \mathcal{N}\giventhat*{\boldsymbol{y}_{\cdot j}}{\mathbf{L}\boldsymbol{f}_{\cdot j},\text{diag}{\left(\left\{\tau_i^{-1}\right\}_{i=1}^G\right)}},
\end{equation}
where $\boldsymbol{y}_{\cdot j}$ and $\boldsymbol{f}_{\cdot j}$ indicate the $j$-th column of $\mathbf{Y}$ and $\mathbf{F}$ respectively, and $\tau_i$ is the precision of normal errors for observations on feature~$i$.

\textbf{Prior specifications.} To induce sparsity in the loading matrix $\mathbf{L}$, we introduce a binary matrix $\mathbf{Z}\in \mathbb{R}^{G\times K}$ whose entries are 1 when the corresponding loading is nonzero. We then specify the following spike-and-slab prior:
\begin{equation}
    p\giventhat*{l_{ik}}{z_{ik}, \alpha_k} = 
    \begin{cases}
        \delta_0{\left(l_{ik}\right)} &\text{if }z_{ik} = 0\\
        \mathcal{N}\giventhat*{l_{ik}}{0, \alpha_k^{-1}} &\text{if }z_{ik} = 1
    \end{cases},
\end{equation}
where $\delta_0$ is the Dirac delta distribution, $l_{ik}$ is the loading of factor $k$ on feature $i$, $z_{ik}$ is a binary variable which indicates whether feature~$i$ is related to factor~$k$, and $\alpha_k$ is the factor-specific normal precision of the nonzero values of $l_{ik}$. Independent Bernoulli priors are placed on the connectivity matrix $\mathbf{Z}$:
\begin{equation}
    p{\left(z_{ik}\right)} = \text{Bernoulli}\giventhat*{z_{ik}}{\pi_{k}},
\end{equation}
where $\boldsymbol{\pi} = \left\{\pi_{k}\right\}_{k=1}^K$ are hyperparameters to be specified. Note that $\pi_{k}$ controls the sparsity of column $k$ of $\mathbf{Z}$, which corresponds to factor $k$.
A gamma prior (shape-rate parametrisation) is imposed on the precisions of the loading matrix $\mathbf{L}$:
\begin{equation}
    p{\left(\alpha_k\right)} = \Gamma\giventhat*{\alpha_k}{a_\alpha,b_\alpha},
\end{equation}
where $a_\alpha$ and $b_\alpha$ are hyperparameters to be specified.

To avoid non-identifiability issues caused by scaling \citep{fatf, manu}, a unit variance normal prior is used for the activation matrix $\mathbf{F}$:
\begin{equation}
    p{\left(\boldsymbol{f}_{\cdot j}\right)} = \mathcal{N}\giventhat*{\boldsymbol{f}_{\cdot j}}{\mathbf{0},\mathbf{I}},
\end{equation}
where $\mathbf{I}$ is the identity matrix of size $K$. Lastly, a gamma prior is placed on the precision parameters of the error model:
\begin{equation}
    p{\left(\tau_i\right)} = \Gamma\giventhat*{\tau_i}{a_\tau,b_\tau},
\end{equation}
where $a_\tau$ and $b_\tau$ are hyperparameters to be specified.

\textbf{Bayesian inference.} Bayesian inference aims to find the posterior distribution $p\giventhat*{\mathbf{L}, \mathbf{F}, \mathbf{Z}, \boldsymbol{\tau}, \boldsymbol{\alpha}}{\mathbf{Y}}$. An exact calculation of the posterior distribution is intractable, so we resort to approximate methods to obtain the posterior distribution. The next two sections describe two possible Bayesian inference techniques for the sparse factor model, namely Markov chain Monte Carlo and variational inference. 

\section{Markov chain Monte Carlo}
\emph{Markov chain Monte Carlo} (\textsc{mcmc}) is a family of algorithms which simulate the posterior distribution $p{(\boldsymbol\theta| \mathbf{Y})}$, where $\boldsymbol\theta$ and $\mathbf{Y}$ denote model parameters and data respectively. In particular, \textsc{mcmc} simulates samples from $p{(\boldsymbol\theta| \mathbf{Y})}$ by constructing a Markov chain $\left\{\boldsymbol\theta^{(n)}\right\}_{n=1}$ that converges to $p{(\boldsymbol\theta| \mathbf{Y})}$. \emph{Gibbs sampler} is a \textsc{mcmc} sampler for a multivariate $\boldsymbol\theta = (\theta_1,\ldots, \theta_m)$ which uses full conditional distributions to construct the Markov chain. Specifically, the transition probability of the chain (assuming a fixed ordering) can be written as
\begin{equation}
    p\giventhat*{\boldsymbol\theta^{(n)}}{\boldsymbol\theta^{(n-1)}} = \prod_{i=1}^m p\giventhat*{\theta^{(n)}_i}{
    \theta^{(n)}_1,
   	\ldots,
    \theta^{(n)}_{i - 1},
    \theta^{(n-1)}_{i + 1},
    \ldots,
    \theta^{(n-1)}_{m}, \mathbf{Y}}.
\end{equation}
That is, the Gibbs sampler cycles through sampling each parameter (or parameter block) from its full conditional posterior distribution. 

\textbf{Collapsed Gibbs sampler for the sparse factor model.} In the sparse factor model, there is a strong dependence between the parameters $l_{ik}$ and $z_{ik}$, as they must be either both zero or both nonzero. Hence, applying standard Gibbs sampler to the sparse factor model will lead to slow mixing. To improve the mixing of the chain, a collapsed Gibbs sampler is used, following the approach of \cite{manu}. Specifically, $\mathbf{L}$ is marginalised out from the conditional distribution of $\mathbf{Z}$, so that $z_{ik}$ is sampled from $p\giventhat*{z_{ik}}{\mathbf{Y}, \mathbf{F}, \mathbf{Z}_{-ik}, \boldsymbol{\tau}, \boldsymbol{\alpha}}$ instead of the full conditional $p\giventhat*{z_{ik}}{\mathbf{Y}, \mathbf{L}, \mathbf{F}, \mathbf{Z}_{-ik}, \boldsymbol{\tau}, \boldsymbol{\alpha}}$, where $\mathbf{Z}_{-ik}$ denotes the elements of $\mathbf{Z}$ excluding $z_{ik}$. Algorithm~\ref{cgs} describes this sampler in full, and the derivations of the conditional distributions can be found in Appendix~\ref{dg}.

\begin{algorithm}[h]
\KwIn{$T, \mathbf{Y}, \boldsymbol{\pi}, a_\tau, b_\tau, a_\alpha, b_\alpha$}
\KwOut{$T$ samples approximating the posterior distribution}
randomly initialise $\mathbf{L}', \mathbf{F}', \mathbf{Z}', \boldsymbol{\tau}', \boldsymbol{\alpha}'$ (most recent sample)\; 
\For{$t\leftarrow 1$ \KwTo $T$}{
    \For{$i\leftarrow 1$ \KwTo $G$}{
        \For{$k\leftarrow 1$ \KwTo $K$}{
            $z'_{ik}\leftarrow z_{ik}^{(t)}\sim p\giventhat*{z_{ik}}{\mathbf{Y}, \mathbf{F}', \mathbf{Z}'_{-ik}, \boldsymbol{\tau}', \boldsymbol{\alpha}', \boldsymbol{\pi}}$\;
        }
    }
    $\mathbf{L}'\leftarrow \mathbf{L}^{(t)}\sim p\giventhat*{\mathbf{L}}{\mathbf{Y}, \mathbf{F}', \mathbf{Z}', \boldsymbol{\tau}', \boldsymbol{\alpha}'}$\;
    $\mathbf{F}'\leftarrow \mathbf{F}^{(t)}\sim p\giventhat*{\mathbf{F}}{\mathbf{Y}, \mathbf{L}', \mathbf{Z}', \boldsymbol{\tau}', \boldsymbol{\alpha}'}$\;
    $\boldsymbol{\tau}'\leftarrow \boldsymbol{\tau}^{(t)}\sim p\giventhat*{\boldsymbol{\tau}}{\mathbf{Y}, \mathbf{L}', \mathbf{F}', \mathbf{Z}', \boldsymbol{\alpha}', a_\tau, b_\tau}$\;
    $\boldsymbol{\alpha}'\leftarrow \boldsymbol{\alpha}^{(t)}\sim p\giventhat*{\boldsymbol{\alpha}}{\mathbf{Y}, \mathbf{L}', \mathbf{F}', \mathbf{Z}', \boldsymbol{\tau}', a_\alpha, b_\alpha}$\;
}
\KwRet{$\left\{\mathbf{L}^{(t)}, \mathbf{F}^{(t)}, \mathbf{Z}^{(t)}, \boldsymbol{\tau}^{(t)}, \boldsymbol{\alpha}^{(t)}\right\}_{t=1}^T$}
\caption{Collapsed Gibbs sampler for the sparse factor model}\label{cgs}
\end{algorithm}

\textbf{Handling the symmetry of the sparse factor model.}

Given a mode of the posterior distribution, if factors (of equal $\pi_k$) are permuted, or if the sign of the entries of $\mathbf{L}$ and $\mathbf{F}$ corresponding to a factor are switched, one obtains another equivalent mode. These symmetries result in up to $2^K K!$ equivalent modes in the posterior distribution, implying that the model is non-identifiable. A \textsc{mcmc} sampler for this model potentially suffers from the label-switching or sign-switching issue. If this occurs, posterior averages will not provide meaningful summaries of the information available; see \citet{relabel} for more discussion. 

For our Gibbs sampler, label-switching or sign-switching rarely happens within a chain, and each chain usually explores only one of the equivalent modes. This is because we simulate $\mathbf{L}$ and $\mathbf{F}$ in separate steps. For example, when sampling $\mathbf{L}$, it is unlikely to have one of its column's signs flipped (for large enough $G$) while $\mathbf{F}$ is held constant. Similar behaviour of \textsc{mcmc} samplers has been previously noted in \citet{structure}. Exploring a single mode corresponding to a particular labelling of the factors is not a huge problem because the equivalent modes from permuted factors are the same from the point of view of inferring a set of factors. The ambiguity of the sign could be resolved later based on domain-specific knowledge, such as genes known to be up-regulated in a particular pathway~\citep{manu}.


Nevertheless, model non-identifiability is still an issue when it is desired to combine multiple chains from different starting values as each chain may explore a different mode. Thus, we implemented a relabelling algorithm \cite{relabel} to deal with this issue, as well as any potential label-switching or sign-switching issue during sampling. See Appendix~\ref{rs} for details of our relabelling algorithm.


\section{Variational inference}
\emph{Variational inference} (\textsc{vi}) is a method from machine learning that approximates probability distributions using optimisation \citep{viblei}, serving as an alternative approach to \textsc{mcmc}. We first review \textsc{vi} in Section~\ref{subsec:vi}, and then describe its application to the sparse factor model in Section~\ref{subsec:vi_sfa}. Further background on \textsc{vi} can be found in \citet{viblei}.

\subsection{Variational inference as a Bayesian inference technique} \label{subsec:vi}
Let $\boldsymbol\theta$ and $\mathbf{Y}$ denote the model parameters and data, respectively. Instead of sampling from the posterior distribution $p\giventhat*{\boldsymbol\theta}{\mathbf{Y}}$, \textsc{vi} approximates the posterior distribution by recasting the inference problem into an optimisation problem. Given a family of probability distributions $\mathcal{D}$, \textsc{vi} aims to find the member of $\mathcal{D}$ (called the variational approximation) which minimises its Kullback-Leibler (\textsc{kl}) divergence to the exact posterior,
\begin{equation}
\label{eq:vi_op}
    q^*{(\boldsymbol\theta)}
    = \argmin_{q{(\boldsymbol\theta)}\in\mathcal{D}} \kl*{ q{(\boldsymbol\theta)} }{ p\giventhat*{\boldsymbol\theta}{\mathbf{Y}} }
    = \argmin_{q{(\boldsymbol\theta)}\in\mathcal{D}} \E{\log q{(\boldsymbol\theta)} - \log p\giventhat*{\boldsymbol\theta}{\mathbf{Y}}},
\end{equation}
where the expectation is taken with respect to $q$. \textsc{kl} divergence penalises choices of $q$ which place significant probability mass on areas where $p$ has little probability mass, thus coercing the density of $q$ to match that of $p$. It does not however, penalise as much the choices of $q$ which place less probability mass on areas where $p$ has more probability mass. Therefore, \textsc{vi} tends to underestimate the variance of the posterior distribution \citep{viblei}. Another implication of this penalisation is that \textsc{vi} attempts to match the most significant modes of $p$ and $q$, potentially disregarding other modes of $p$ that are further away. Thus, when applied to the sparse factor model, $q$ tends to capture only one of the modes. Finally, a choice of $\mathcal{D}$ that is too restrictive may result in a variational approximation that does not capture the posterior distribution accurately.

\textbf{Evidence lower bound.} In practice, the \textsc{kl} divergence cannot be computed directly, but is related to the \emph{evidence lower bound} $\textsc{elbo}{(q)} = \E{\log p{(\mathbf{Y}, \boldsymbol\theta)} - \log q{(\boldsymbol\theta)}}$ by the equation
\begin{equation}
    \kl*{ q{(\boldsymbol\theta)} }{ p\giventhat*{\boldsymbol\theta}{\mathbf{Y}} }
    = \E{\log q{(\boldsymbol\theta)} - \log p{(\mathbf{Y}, \boldsymbol\theta)} + \log p{(\mathbf{Y})}}
    = -\textsc{elbo}{(q)} + \log p{(\mathbf{Y})}.
\end{equation}
Since $\log p{(\mathbf{Y})}$ is constant, minimising the \textsc{kl} divergence is equivalent to maximising the \textsc{elbo}. As the \textsc{kl} divergence is always nonnegative \citep{kl}, it follows  that $\textsc{elbo}{(q)}\le \log p{(\mathbf{Y})}$, hence the name evidence lower bound. Provided that the family of distributions $\mathcal{D}$ is simple enough, the \textsc{elbo} is a tractable quantity to compute.

\textbf{Mean-field approximation and coordinate ascent variational inference.} A common choice of $\mathcal{D}$ is the \emph{mean-field variational family}, where the model parameters $\boldsymbol\theta = \left\{\theta_i\right\}^{m}_{i=1}$ are mutually independent in $q$. In other words, the variational approximation can be written as a product of variational factors,
\begin{equation}
    q{(\boldsymbol\theta)} = \prod_{i=1}^m q_i{(\theta_i)}.
\end{equation}
One of the most commonly used algorithms for solving the optimisation problem in equation~\eqref{eq:vi_op} with the mean-field family is coordinate ascent variational inference (\textsc{cavi}) \citep{bishop}. The \textsc{cavi} algorithm iterates through the variational factors, updating each $q_i{(\theta_i)}$ while holding the other variational factors fixed:
\begin{equation}
\label{eq:cavi}
    q_i^*{(\theta_i)} \propto
    \exp\left\{\Eover{-i}{\log p\giventhat*{\theta_i}{\mathbf{Y},\boldsymbol\theta_{-i}}}\right\} \propto
    \exp\left\{\Eover{-i}{\log p{\left(\mathbf{Y},\boldsymbol\theta\right)}}\right\},
\end{equation}
where the expectation $\mathbb{E}_{-i}$ is taken with respect to the currently fixed variational factors, $\prod_{j\neq i}^m q_j{(\theta_j)}$. This update maximises the \textsc{elbo} given the currently fixed variational factors \citep{viblei}. This enables the algorithm to monotonically optimise the \textsc{elbo}, eventually reaching a local optimum.  


\subsection{Variational inference for the sparse factor model} \label{subsec:vi_sfa}
For the sparse factor model, we choose the following mean-field variational family to approximate the posterior distribution:
\begin{equation}
    q{\left(\mathbf{L},\mathbf{F},\mathbf{Z},\boldsymbol{\tau},\boldsymbol\alpha\right)}
    = \prod_{i=1}^{G} \left[ q{(\tau_i)} \prod_{k=1}^K q{\left(l_{ik},z_{ik}\right)} \right]
    \times \prod_{k=1}^K \left[ q{\left(\alpha_k\right)} \prod_{j=1}^{N} q{\left(f_{kj}\right)} \right],  
\end{equation}
where
\begin{align}
    q{\left(l_{ik},z_{ik}\right)}
    &= q{\giventhat*{l_{ik}}{z_{ik}}}q{\left(z_{ik}\right)}\\ &= \mathcal{N}\giventhat*{l_{ik}}{\mu_{l_{ik}},\sigma^2_{l_{ik}}}^{z_{ik}}
    \times\delta_0{\left(l_{ik}\right)}^{1-z_{ik}}\times \text{Bernoulli}\giventhat*{z_{ik}}{\eta_{ik}}\\
    q{\left(f_{kj}\right)} &=
    \mathcal{N}\giventhat*{f_{kj}}{\mu_{f_{kj}},\sigma^2_{f_{kj}}} \\
    q{\left(\tau_i\right)} &=
    \Gamma\giventhat*{\tau_i}{\hat{a}_{\tau_i},\hat{b}_{\tau_i}} \\
    q{\left(\alpha_k\right)} &=
    \Gamma\giventhat*{\alpha_k}{\hat{a}_{\alpha_k},\hat{b}_{\alpha_k}}.
\end{align}
Each variational factor we choose is conjugate to the distribution in the likelihood function, so the variational family satisfies the update rule in equation~\eqref{eq:cavi}, and an analytic computation of the expectation on the right is possible. We use \textsc{cavi} to optimise the \textsc{elbo}. Algorithm~\ref{cavi} shows our \textsc{cavi} for the sparse factor model; see Appendix~\ref{dc} for details of the \textsc{cavi} updates and the derivation of the \textsc{elbo}.

Note that the variational factor $q{\left(l_{ik},z_{ik}\right)}$ does not factorise into $q{\left(l_{ik}\right)}q{\left(z_{ik}\right)}$, as it is not possible to remove the dependency of $l_{ik}$ and $z_{ik}$ being either both zero or both nonzero. Moreover, we derived the variational factor for the exact spike and slab prior instead of the relaxed sparsity prior used in \citet{manu}.

\begin{algorithm}[h]
\KwIn{$\mathbf{Y}, \boldsymbol{\pi}, a_\tau, b_\tau, a_\alpha, b_\alpha$}
\KwOut{variational factors which approximate the posterior distribution}
randomly initialise $
q{\left(l_{ik},z_{ik}\right)},
q{\left(f_{kj}\right)},
q{\left(\tau_i\right)},
q{\left(\alpha_k\right)}
\;\forall i,j,k$\;
\While{\textsc{elbo} has not converged}{
    \For{$i\leftarrow 1$ \KwTo $G$}{
        \For{$k\leftarrow 1$ \KwTo $K$}{
            $q{\left(l_{ik},z_{ik}\right)}
    \propto \exp{\left\{\Eover{\mathbf{L}_{-ik},\mathbf{F},\mathbf{Z}_{-ik},\tau_i,\boldsymbol{\alpha}}{\log p\giventhat*{l_{ik},z_{ik}}{\mathbf{Y},\mathbf{L}_{-ik},\mathbf{F},\mathbf{Z}_{-ik},\boldsymbol{\tau},\boldsymbol\alpha}}\right\}}$\;
        }
    }
    \For{$k\leftarrow 1$ \KwTo $K$}{
        \For{$j\leftarrow 1$ \KwTo $N$}{
            $q{\left(f_{kj}\right)}
    \propto \exp{\left\{\Eover{\mathbf{L},\mathbf{F}_{-kj},\mathbf{Z},\boldsymbol{\tau},\boldsymbol\alpha}{\log p\giventhat*{f_{kj}}{\mathbf{Y},\mathbf{L},\mathbf{F}_{-kj},\mathbf{Z},\boldsymbol{\tau},\boldsymbol\alpha}}\right\}}$\;
        }
    }
    \For{$i\leftarrow 1$ \KwTo $G$}{
        $q{\left(\tau_i\right)}
    \propto
    \exp{\left\{\Eover{\mathbf{L},\mathbf{F},\mathbf{Z},\boldsymbol\alpha}{\log p\giventhat*{\tau_i}{\mathbf{Y},\mathbf{L},\mathbf{F},\mathbf{Z},\boldsymbol\alpha}}\right\}}$\;
    }
    \For{$k\leftarrow 1$ \KwTo $K$}{
        $q{\left(\alpha_k\right)}
    \propto
    \exp{\left\{\Eover{\mathbf{L},\mathbf{F},\mathbf{Z},\boldsymbol\tau}{\log p\giventhat*{\alpha_k}{\mathbf{Y},\mathbf{L},\mathbf{F},\mathbf{Z},\boldsymbol\tau}}\right\}}$\;
    }
}
\KwRet{$
q{\left(l_{ik},z_{ik}\right)},
q{\left(f_{kj}\right)},
q{\left(\tau_i\right)},
q{\left(\alpha_k\right)}
\;\forall i,j,k$}
\caption{\textsc{cavi} for the sparse factor model}\label{cavi}
\end{algorithm}

\textbf{Initialisation.} \textsc{cavi} is a hill-climbing algorithm that may find only a local optimum of the \textsc{elbo}. In practice, we run multiple \textsc{vi} trials with different initialisations, and select the trial that converges to the largest \textsc{elbo} for inference. To reduce computation, trials may be stopped early, and only the trial corresponding to the largest \textsc{elbo} (at early stopping) is run until convergence.

\section{Numerical comparisons}

We compare the performance of \textsc{mcmc} and \textsc{vi}, focusing on accuracy and computational efficiency. It is expected that \textsc{vi} would converge faster than \textsc{mcmc}, but \textsc{mcmc} will provide more accurate inference in the long run. The comparison is carried out for simulated datasets and a real biological dataset.

\subsection{Simulated data}
\begin{figure*}[t]
    \centering
    \includegraphics[width=6in]{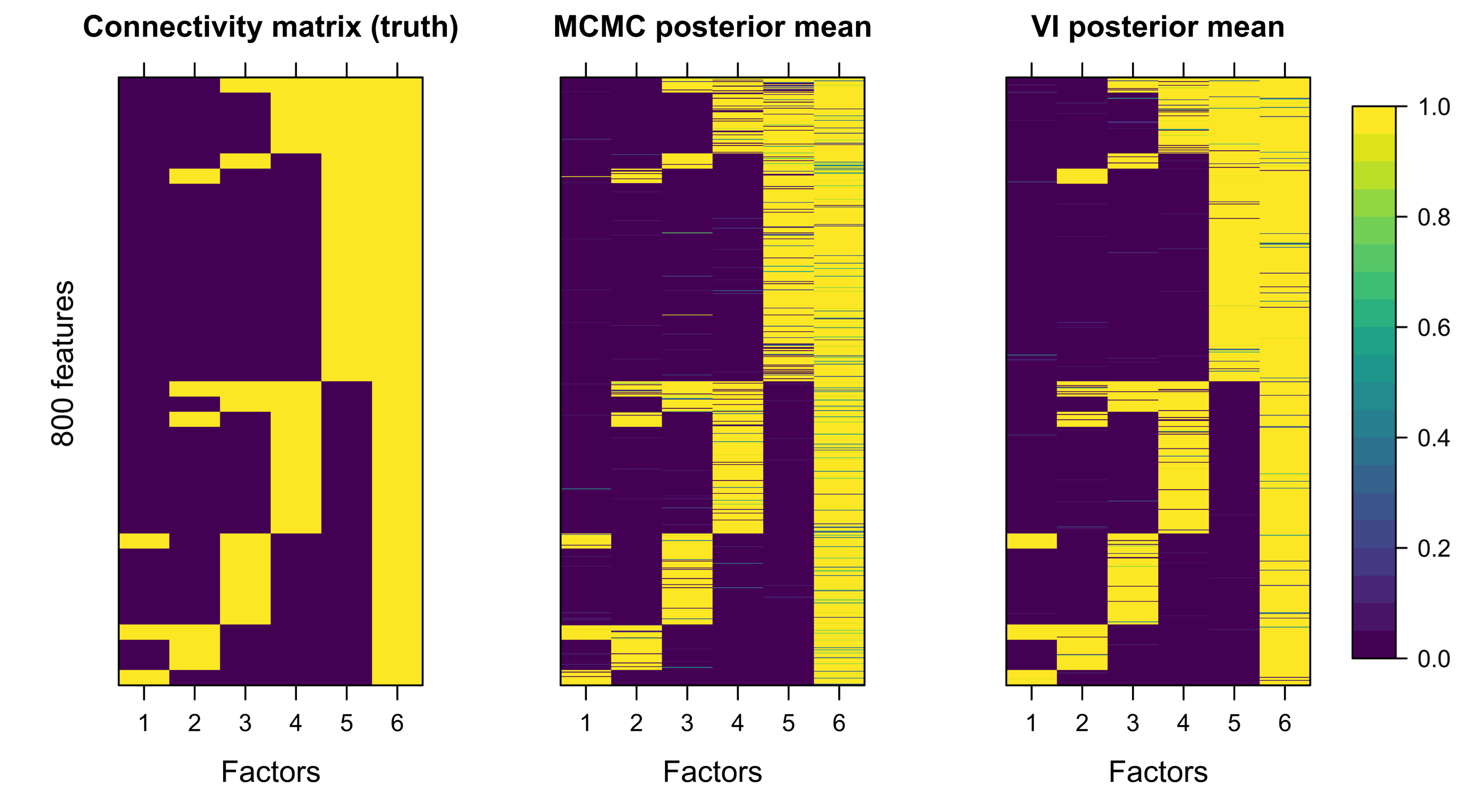}
    \caption{True connectivity structure for $\mathbf{Z}$ (simulated with snr = 5), and inferred structures (posterior mean of $\mathbf{Z}$). Results from a \textsc{mcmc} chain with the best accuracy of $\mathbf{Z}$ and a \textsc{vi} trial with the largest converged \textsc{elbo} are shown.}\label{sim_zmat}
\end{figure*}

The simulated datasets each consist of $G$~=~800 features over $N$~=~100 samples, explained by $K$~=~6 factors. We simulated three datasets with varying amount of noise to evaluate the robustness of each inference technique. All three datasets share the same underlying connectivity structure $\mathbf{Z}$ (the first panel of Figure~\ref{sim_zmat}), consisting of 5 factors with sparse loadings and 1 factor with full loadings, corresponding to the sparsity hyperparameters $\boldsymbol\pi$~=~(0.075, 0.15, 0.25, 0.375, 0.5, 1). The entries of $\mathbf{L}$ (that correspond to $z_{ik} = 1$) and $\mathbf{F}$ were simulated from independent standard normal distributions. The random errors present in each dataset was controlled by varying the signal-to-noise ratio (snr~=~1, 5, 25). We quantified the signal for feature~$i$ using the sample variance $V_i$ of the entries in row $i$ of $\mathbf{LF}$ (the expectation of the data for feature $i$). The precision of the error is then given by
\begin{equation}
    \tau_i = \frac{\text{snr}}{V_i}.
\end{equation}

\begin{figure*}[t]
    \centering
    \includegraphics[width=5in]{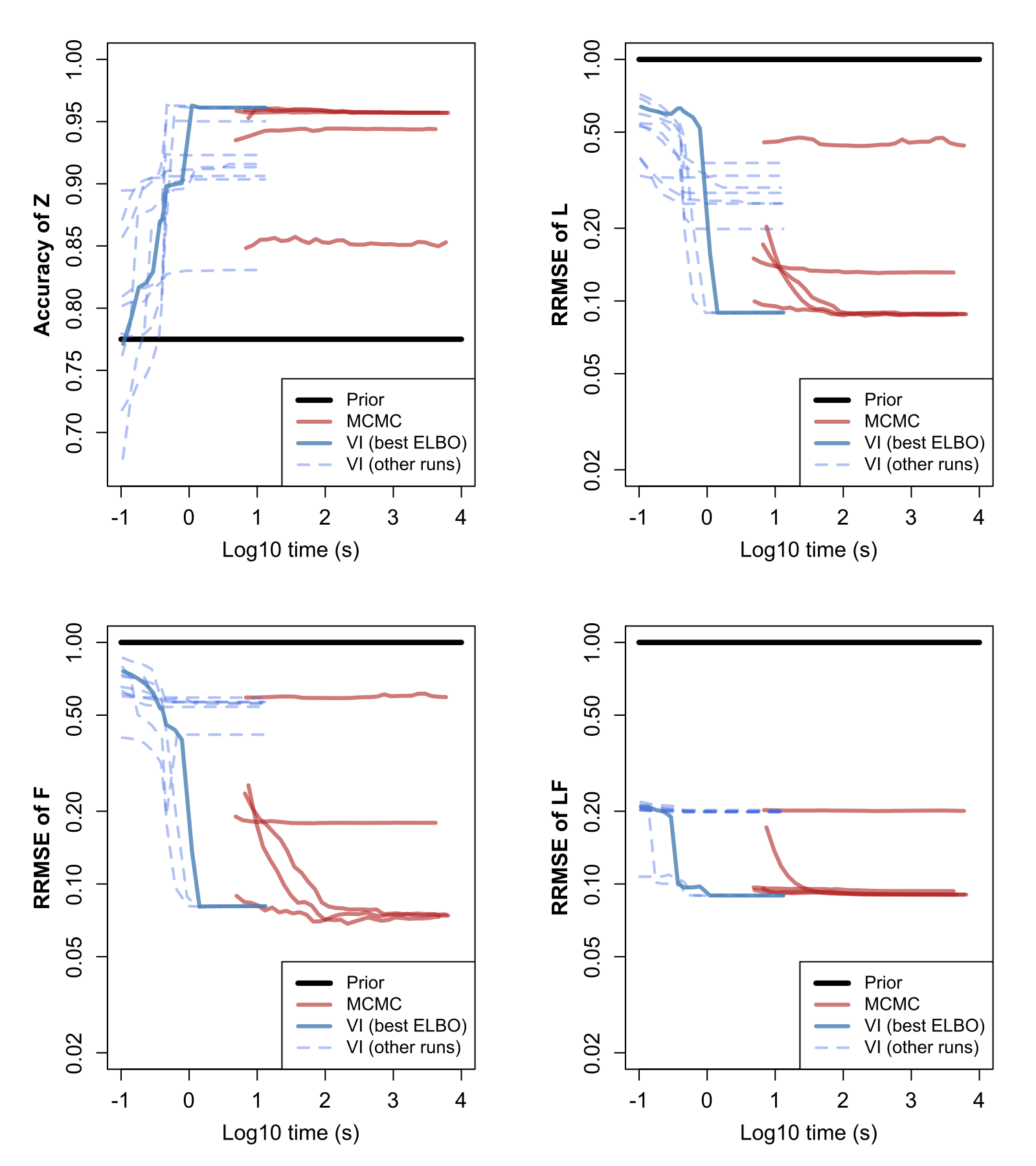}
    \caption{Performance over computation time on a simulated dataset with a moderate amount of noise (snr = 5), based on the posterior mean of the connectivity structure $\mathbf{Z}$, loading matrix $\mathbf{L}$, activation matrix $\mathbf{F}$, and low-dimensional structure $\mathbf{LF}$.}\label{sim_perf_timing}
\end{figure*}

We applied $\textsc{mcmc}$ and \textsc{vi} to each of these datasets, assuming \emph{a priori} that we know the correct number of sparse factors and dense factors (5 and 1 respectively). The sparsity hyperparameters $\boldsymbol\pi$ were set to be 0.1 and 0.9 for sparse factors and dense factors respectively. The remaining hyperparameters for the gamma priors were set to be $a_\tau = b_\tau = a_\alpha = b_\alpha = 10^{-3}$, corresponding to vague priors. For $\textsc{mcmc}$, we discarded the first 100 iterations as a burn-in, and then ran 200,000 iterations, keeping one out of every 10 successive samples for inference. We ran $\textsc{mcmc}$ 5 times with different initial values, giving 5 chains of 20,000 samples each. We ran 10 \textsc{vi} trials until the \textsc{elbo} converged (up to absolute difference of $10^{-10}$ or relative difference of $10^{-14}$).



\begin{figure*}[p]
    \centering
  \includegraphics[width=5.1in]{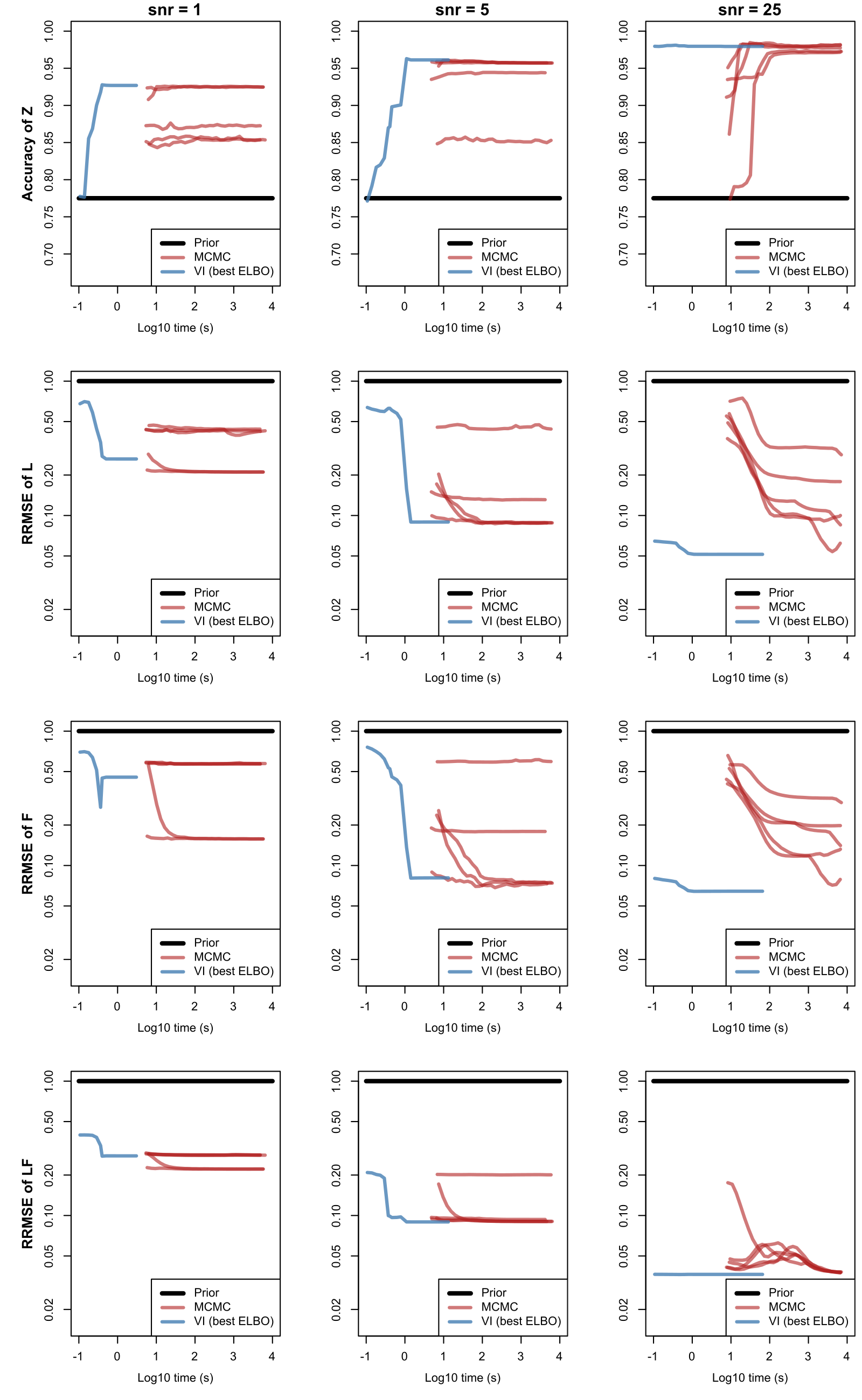}
    \caption{Performance over computation time across different simulated datasets with varying amounts of noise (snr = 1, 5, 25), based on the posterior mean of the connectivity structure $\mathbf{Z}$, loading matrix $\mathbf{L}$, activation matrix $\mathbf{F}$, and low-dimensional structure $\mathbf{LF}$.}\label{sim_robust}
\end{figure*}

\textbf{Comparison of accuracy and speed.} We first compare the performance of \textsc{mcmc} and \textsc{vi} using a dataset with a moderate amount of noise (snr = 5). Performance is evaluated via the accuracy of the inferred connectivity structure $\mathbf{Z}$, loading matrix $\mathbf{L}$, activation matrix $\mathbf{F}$, and low-dimensional structure $\mathbf{LF}$. The accuracy of $\mathbf{Z}$ is defined as the proportion of correctly inferred entries after rounding the posterior means of $\mathbf{Z}$ to 0 or 1. The accuracy of $\mathbf{L}$, $\mathbf{F}$, and $\mathbf{LF}$ is quantified by the relative root mean squared error (\textsc{rrmse}). As an example, the \textsc{rrmse} for $\mathbf{L}$ is
\begin{equation}
    \textsc{rrmse}{(\hat{\mathbf{L}}, \mathbf{L})} = \sqrt{\frac{\sum_{i,k}(\hat{l}_{ik} - l_{ik})^2}{\sum_{i,k} l_{ik}^2}},
\end{equation}
where $\hat{\mathbf{L}}$ is the posterior mean of $\mathbf{L}$. 
We included the performance measures of the prior mean as a baseline to compare to. These performance measures were calculated after the inferred model parameters have been permuted and scaled appropriately to match the simulation parameters. 

Figure~\ref{sim_perf_timing} shows the accuracy of each method over computation time. Three out of the five \textsc{mcmc} chains captured the underlying structure well, as evident from a high accuracy of $\mathbf{Z}$ and small \textsc{rrmse} of $\mathbf{L}$, $\mathbf{F}$ and $\mathbf{LF}$. Two chains failed to converge, even after more than two hours of running 200,000 \textsc{mcmc} iterations for each chain. In contrast, all \textsc{vi} trials converged in about 10 seconds, although the performance varied across trials. This is expected, as each trial climbs the \textsc{elbo} to a different local optimum. Moreover, the trial which converged to the largest \textsc{elbo} does not display any significant loss in accuracy when compared to the best accuracy achieved by \textsc{mcmc}.

Figure~\ref{sim_zmat} (the second and third panels) presents a visualisation of the inferred connectivity structure $\mathbf{Z}$ from the MCMC chain with the best accuracy and the VI trial with the largest ELBO, showing that both techniques are capable of discovering $\mathbf{Z}$. False negatives observed in the results from both methods most likely correspond to small factor loadings that were shrunk to zero. 

\textbf{Robustness against noise.} Now we compare the performance of \textsc{mcmc} and \textsc{vi} when applied to datasets with different amounts of noise (snr =1, 5, 25). The best \textsc{vi} trial (best in the sense of largest converged \textsc{elbo}) achieved better performance as the amount of noise decreases (Figure~\ref{sim_robust}). In all cases, its accuracy of $\mathbf{Z}$ roughly matched that of the most accurate result from a \textsc{mcmc} chain. The only case where \textsc{mcmc} may have an advantage is the dataset with snr = 1, where 2 out of the 5 \textsc{mcmc} chains achieved a lower error on $\mathbf{L}$, $\mathbf{F}$, and $\mathbf{LF}$ than the best \textsc{vi} trial. In fact, these 2 chains managed to accurately infer factor 1, which is the factor with the most sparsity. The remaining 3 chains and all 10 \textsc{vi} trials did not find this factor. \textsc{mcmc} may be more capable to infer sparse factors from noisy data than \textsc{vi}, but does not do so consistently. 

The accuracy measures for \textsc{mcmc} took a longer time to converge for the dataset with the least noise (snr = 25). A possible explanation is that stronger signals make the dependency structure in the posterior distribution stronger, leading to less efficient convergence of the Gibbs sampler. In the noisier datasets, some \textsc{mcmc} chains were clearly stuck in non-optimal modes that do not match the underlying structure.

\subsection{Biological data}
In this section, we compare the performance of \textsc{mcmc} and \textsc{vi} when applying the sparse factor model to a real dataset. To this end, we used {\it GTEx eQTL summary data} from \cite{ebmf}, which consists of $Z$-scores measuring the associations of $G$ = 16069 genetic variants with gene expression measured in $N$ = 44 human tissues. In other words, $y_{ij}$ indicates the strength of effect of genetic variant~$i$ on gene expression in tissue~$j$. This dataset originates from the Genotype Tissue Expression (GTEx) Project \citep{gtex}, which \cite{ebmf} used as part of their evaluation of \emph{flash}, a \textsc{vi}-based method they developed for an empirical Bayes approach to matrix factorisation.  See \cite{ebmf} for a further description of the GTEx eQTL summary data. 

The \emph{flash} is capable of automatically selecting the number of factors $K$, which \cite{ebmf} report to be $K$ = 26 when applied to this dataset. We used the same number of factors as inferred by \emph{flash}, and treated all 26 factors as sparse factors, each with a sparsity hyperparameter of $\pi_k = 0.1$. The remaining hyperparameters for the gamma priors remained at $10^{-3}$. The first 2,000 $\textsc{mcmc}$ iterations were discarded as a burn-in. After the burn-in period, 16,000 iterations were run, where one out of every 10 successive samples were kept for inference. We ran $\textsc{mcmc}$ 5 times with different starting points, giving 5 chains of 1,600 samples each. We ran 10 \textsc{vi} trials until the \textsc{elbo} converged up to a tolerance of $10^{-3}$.

\begin{figure*}[t]
    \centering
    \includegraphics[width=3in]{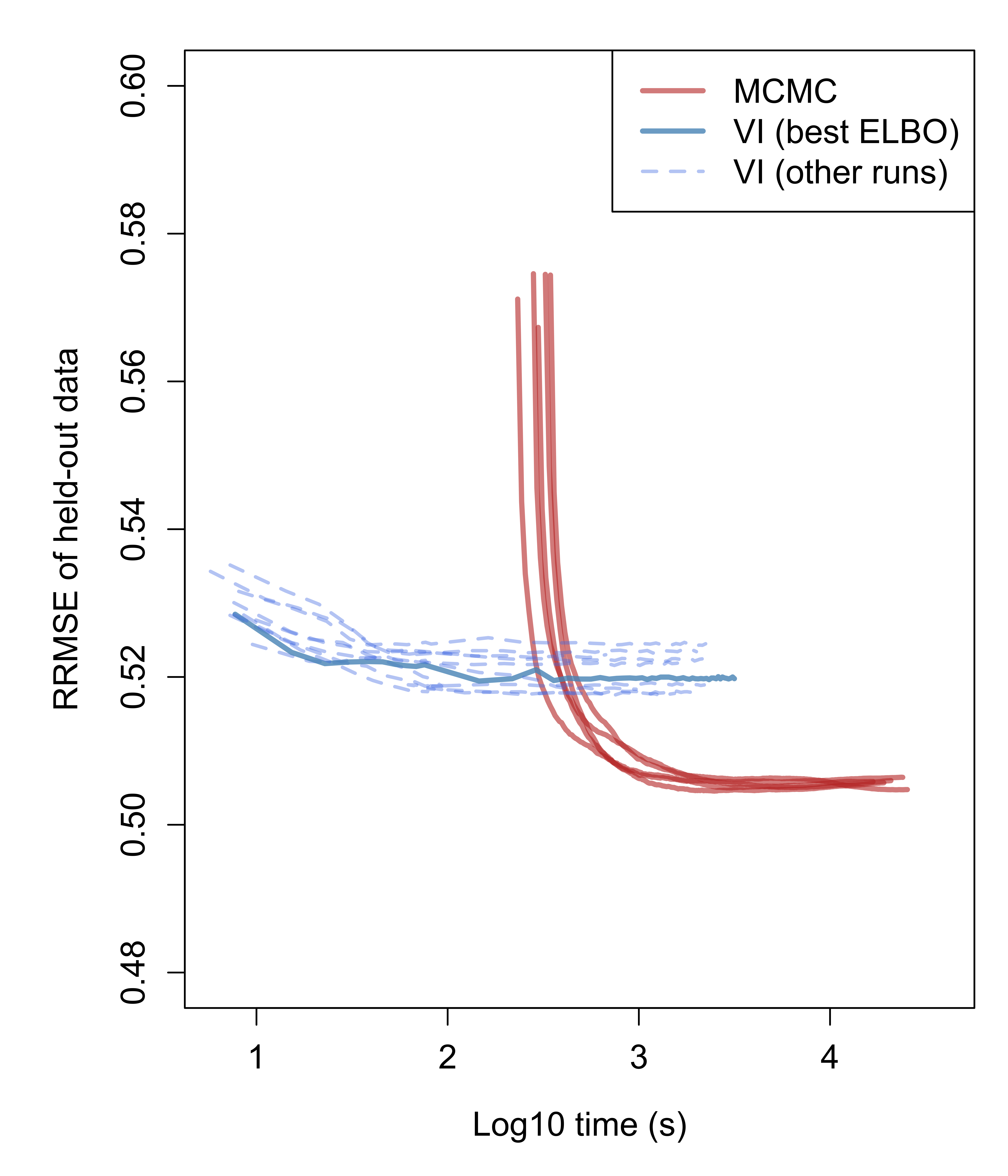}
    \caption{Performance on GTEx eQTL summary data over computation time, based on the posterior mean of predictions on held-out data (10\% of full data).}\label{gtex_perf}
\end{figure*}

\textbf{Fill-in test.} As the ground truth for an underlying structure is not available, we assessed the performance of each method using a \emph{fill-in test}, following \cite{manu} and \cite{ebmf}. We first held-out (masked) 70704 data entries in $\mathbf{Y}$, corresponding to 10\% of the data entries. Then, we inferred the model parameters using the remaining 90\% of the data and predicted (filled-in) these 70704 missing values using the inferred parameters. Finally, we assessed the performance of each method using the RRMSE of the posterior mean of predictions on the held-out data, against the observed held-out data. The idea is that model parameters which better capture the true underlying structure will predict the held-out entries more accurately \citep{manu}. As expected, \textsc{vi} is computationally more efficient than \textsc{mcmc} (Figure~\ref{gtex_perf}), and its RRMSE is only slightly worse than that of \textsc{mcmc}.

\section{Conclusion}
We have compared two Bayesian inference techniques, \textsc{mcmc} and \textsc{vi}, when applied to the sparse factor model. We have derived and implemented \textsc{mcmc} and \textsc{vi} algorithms, and investigated the relative strengths and weaknesses of two methods in terms of accuracy and computational efficiency using both simulated and biological data. Our empirical investigation showed that \textsc{mcmc} gives more slightly accurate inference than \textsc{vi}, however the difference is outweighed by the much faster speed of \textsc{vi}. After taking into account the need of running multiple \textsc{vi} trials to select the trial with the best \textsc{elbo}, \textsc{vi} achieves similar accuracy as \textsc{mcmc} in significantly less time.


\textbf{Acknowledgements.} Special thanks to Matthew Stephens for sharing the GTEx data used in the numerical comparison and allowing us to make them publicly available. The GTEx Project was supported by the Common Fund of the Office of the Director of the National Institutes of Health, and by NCI, NHGRI, NHLBI, NIDA, NIMH, and NINDS. We thank Yao-ban Chan for helpful comments on a draft manuscript. This research used the Spartan High Performance Computing system at the University of Melbourne. This work was supported by a Vacation Research Scholarship provided by the Australian Mathematical Sciences Institute to Y.S.F.

\textbf{Availability of data and materials.} The GTEx eQTL summary data and our implementation of \textsc{mcmc} and \textsc{vi} algorithms for sparse factor analysis are publicly available at \url{https://github.com/ysfoo/sparsefactor}.

\newpage
\appendixtitleon
\begin{appendices}

\section{Conditional distributions of the sparse factor model}\label{dg}

Denote $D_{\boldsymbol{v}} = \text{diag}{(\boldsymbol{v})}$ for any vector $\boldsymbol{v}$. The full conditional distribution of row~$i$ of $\mathbf{L}$ and $\mathbf{Z}$ is
\begin{align*}
\begin{split}
p\giventhat*{\boldsymbol{l}_{i\cdot}, \boldsymbol{z}_{i\cdot}}{\mathbf{Y},\mathbf{F},\boldsymbol{\tau},\boldsymbol{\alpha}}
\propto {}& \prod_{k\cl z_{ik} = 1} \pi_k\sqrt\frac{\alpha_k}{2\pi}\times \prod_{k\cl z_{ik} = 0} \left(1-\pi_k\right)\delta_0{\left(l_{ik}\right)} \\
&\times \exp\Biggl\{-\frac{\tau_i}{2}\left(\boldsymbol{y}_{i\cdot}-\left[\mathbf{F}\right]^{\mathsf{T}}_{\boldsymbol{z}_{i\cdot}}\left[\boldsymbol{l}_{i\cdot}\right]_{\boldsymbol{z}_{i\cdot}}\right)^{\mathsf{T}}\left(\boldsymbol{y}_{i\cdot}-\left[\mathbf{F}\right]^{\mathsf{T}}_{\boldsymbol{z}_{i\cdot}}\left[\boldsymbol{l}_{i\cdot}\right]_{\boldsymbol{z}_{i\cdot}}\right)\\
&\hspace{4em}-\frac{1}{2}\left[\boldsymbol{l}_{i\cdot}\right]^\mathsf{T}_{\boldsymbol{z}_{i\cdot}}\left[D_{\boldsymbol\alpha}\right]_{\boldsymbol{z}_{i\cdot}}\left[\boldsymbol{l}_{i\cdot}\right]_{\boldsymbol{z}_{i\cdot}}\Biggr\} 
\end{split}\\
\begin{split}
\propto {}& \prod_{k\cl z_{ik} = 1} \pi_k\sqrt\frac{\alpha_k}{2\pi}\times \prod_{k\cl z_{ik} = 0} \left(1-\pi_k\right)\delta_0{\left(l_{ik}\right)} \\
&\times\exp{\left\{-\frac{1}{2}\left(\left[\boldsymbol{l}\right]_{\boldsymbol{z}_{i\cdot}}-\boldsymbol{\mu}_{\boldsymbol{l}_{i\cdot}}\right)^{\mathsf{T}}\Sigma_{\boldsymbol{l}_{i\cdot}}^{-1}\left(\left[\boldsymbol{l}\right]_{\boldsymbol{z}_{i\cdot}}-\boldsymbol{\mu}_{\boldsymbol{l}_{i\cdot}}\right)+\frac{1}{2}\boldsymbol{\mu}_{\boldsymbol{l}_{i\cdot}}^{\mathsf{T}}\Sigma_{\boldsymbol{l}_{i\cdot}}^{-1}\boldsymbol{\mu}_{\boldsymbol{l}_{i\cdot}}\right\}}
\end{split}\label{lzcond}
\end{align*}
where 
\begin{align*}
    \left[\mathbf{F}\right]_{\boldsymbol{z}_{i\cdot}}
    &= \text{matrix consisting of rows of }\mathbf{F}\text{ whose corresponding entries of }\boldsymbol{z}_{i\cdot} \text{ are equal to } 1\\
    \left[\boldsymbol{l}_{i\cdot}\right]_{\boldsymbol{z}_{i\cdot}}
    &= \text{vector consisting of entries of }\boldsymbol{l}_{i\cdot}\text{ whose corresponding entries of }\boldsymbol{z}_{i\cdot} \text{ are equal to } 1\\
    \left[D_{\boldsymbol\alpha}\right]_{\boldsymbol{z}_{i\cdot}}
    &= \text{matrix consisting of rows of }D_{\boldsymbol\alpha}\text{ whose corresponding entries of }\boldsymbol{z}_{i\cdot} \text{ are equal to } 1\\
    \Sigma_{\boldsymbol{l}_{i\cdot}}
    &= \left(\tau_i \left[\mathbf{F}\right]_{\boldsymbol{z}_{i\cdot}}\left[\mathbf{F}\right]_{\boldsymbol{z}_{i\cdot}}^{\mathsf{T}} + \left[D_{\boldsymbol\alpha}\right]_{\boldsymbol{z}_{i\cdot}}\right)^{-1}\\
    \boldsymbol{\mu}_{\boldsymbol{l}_{i\cdot}}
    &= \tau_i\Sigma_{\boldsymbol{l}_{i\cdot}}\left[\mathbf{F}\right]_{\boldsymbol{z}_{i\cdot}}\boldsymbol{y}_{i\cdot}.
\end{align*}
The full conditional distribution of $\boldsymbol{l}_{i\cdot}$ is then
\begin{equation*}
    p\giventhat*{\boldsymbol{l}_{i\cdot}}{\mathbf{Y},\mathbf{F},\mathbf{Z},\boldsymbol{\tau},\boldsymbol\alpha} = \mathcal{N}\giventhat*{\left[\boldsymbol{l}_{i\cdot}\right]_{\boldsymbol{z}_{i\cdot}}}{\boldsymbol{\mu}_{\boldsymbol{l}_{i\cdot}}, \Sigma_{\boldsymbol{l}_{i\cdot}}}\times \prod_{k\cl z_{ik} = 0} \delta_0{\left(l_{ik}\right)}.
\end{equation*}
To obtain a collapsed Gibbs sampler, $\boldsymbol{l}_{i\cdot}$ is marginalised out from the full conditional distribution of $z_{ik}$:
\begin{equation*}
    p\giventhat*{z_{ik}}{\mathbf{Y},\mathbf{F},\mathbf{Z}_{-ik},\boldsymbol{\tau},\boldsymbol\alpha}
    \propto \left(\frac{\alpha_k}{2\pi}\right)^{\frac{z_{ik}}{2}}\det|\Sigma_{\boldsymbol{l}_{i\cdot}}|^{\frac{1}{2}}\exp{\left\{\frac{1}{2}\boldsymbol{\mu}_{\boldsymbol{l}_{i\cdot}}^{\mathsf{T}}\Sigma_{\boldsymbol{l}_{i\cdot}}^{-1}\boldsymbol{\mu}_{\boldsymbol{l}_{i\cdot}}\right\}}\pi_k^{z_{ik}}\left(1-\pi_k\right)^{1-z_{ik}}.
\end{equation*}
The full conditional distribution of column~$j$ of $\mathbf{F}$ is
\begin{equation*}
    p\giventhat*{\boldsymbol{f}_{\cdot j}}{\mathbf{Y},\mathbf{L},\mathbf{Z},\boldsymbol{\tau},\boldsymbol\alpha}
    \propto \exp{\left\{-\frac{1}{2}\left(\boldsymbol{y}_{\cdot j}-\mathbf{L}\boldsymbol{f}_{\cdot j}\right)^{\mathsf{T}}D_{\boldsymbol{\tau}}\left(\boldsymbol{y}_{\cdot j}-\mathbf{L}\boldsymbol{f}_{\cdot j}\right)-\frac{1}{2}\boldsymbol{f}_{\cdot j}^{\mathsf{T}}\boldsymbol{f}_{\cdot j}\right\}},
\end{equation*}
which is a normal distribution with mean and covariance
\begin{align*}
    \Sigma_{\boldsymbol{f}_{\cdot j}}
    &= \left(\mathbf{L}^{\mathsf{T}}D_{\boldsymbol{\tau}}\mathbf{L} + \mathbf{I}\right)^{-1}\\
    \boldsymbol{\mu}_{\boldsymbol{f}_{\cdot j}}
    &= \Sigma_{\boldsymbol{f}_{\cdot j}}\mathbf{L}^{\mathsf{T}}D_{\boldsymbol{\tau}}\boldsymbol{y}_{\cdot j}.
\end{align*}
Lastly, the full conditional distribution of $\tau_i$ is
\begin{equation*}
    p\giventhat*{\tau_i}{\mathbf{Y},\mathbf{L}, \mathbf{F},\mathbf{Z},\boldsymbol\alpha}
    = \Gamma\giventhat*{\tau_i}{a_\tau + \frac{N}{2},b_\tau + \frac{1}{2}\left(\boldsymbol{y}_{i\cdot}-\left[\mathbf{F}\right]^{\mathsf{T}}_{\boldsymbol{z}_{i\cdot}}\left[\boldsymbol{l}_{i\cdot}\right]_{\boldsymbol{z}_{i\cdot}}\right)^{\mathsf{T}}\left(\boldsymbol{y}_{i\cdot}-\left[\mathbf{F}\right]^{\mathsf{T}}_{\boldsymbol{z}_{i\cdot}}\left[\boldsymbol{l}_{i\cdot}\right]_{\boldsymbol{z}_{i\cdot}}\right)},
\end{equation*}
and the full conditional distribution of $\alpha_k$ is
\begin{equation*}
    p\giventhat*{\alpha_k}{\mathbf{Y},\mathbf{L}, \mathbf{F},\mathbf{Z},\boldsymbol{\tau}} = \Gamma\giventhat*{\alpha_k}{a_\alpha + \frac{1}{2}\sum_{i=1}^G z_{ik},b_\alpha + \frac{1}{2}\sum_{i\cl z_{ik} = 1} l_{ik}^2}.
\end{equation*}

\section{Details of \textsc{vi} for the sparse factor model}\label{dc}

Throughout this section, all expectations are taken over the distribution $q{\left(\mathbf{L},\mathbf{F},\mathbf{Z},\boldsymbol{\tau},\boldsymbol\alpha\right)}$.

\textbf{Coordinate ascent updates.} Coordinate ascent on $\boldsymbol{l}_{i\cdot}$ and $\boldsymbol{z}_{i\cdot}$ gives
\begin{align*}
    q^*{\left(l_{ik},z_{ik}\right)}
    \propto {}&\exp{\left\{\Eover{\mathbf{L}_{-ik},\mathbf{F},\mathbf{Z}_{-ik},\tau_i,\boldsymbol{\alpha}}{\log p\giventhat*{l_{ik},z_{ik}}{\mathbf{Y},\mathbf{L}_{-ik},\mathbf{F},\mathbf{Z}_{-ik},\boldsymbol{\tau},\boldsymbol\alpha}}\right\}}\\
    \begin{split}
    \propto {}&\exp\vast\{ -\Eover{\mathbf{L}_{-ik},\mathbf{F},\mathbf{Z}_{-ik},\tau_i}{\frac{\tau_i}{2}\left(\boldsymbol{y}_{i\cdot}-\left[\mathbf{F}\right]^{\mathsf{T}}_{\boldsymbol{z}_{i\cdot}}\left[\boldsymbol{l}_{i\cdot}\right]_{\boldsymbol{z}_{i\cdot}}\right)^{\mathsf{T}}\left(\boldsymbol{y}_{i\cdot}-\left[\mathbf{F}\right]^{\mathsf{T}}_{\boldsymbol{z}_{i\cdot}}\left[\boldsymbol{l}_{i\cdot}\right]_{\boldsymbol{z}_{i\cdot}}\right)}\\
    &\hspace{3em} +\frac{z_{ik}}{2}\Eover{\alpha_k}{\log\frac{\alpha_k}{2\pi}-\alpha_k l_{ik}^2}\vast\}\times \pi_k^{z_{ik}}\left(\left(1-\pi_k\right)\delta_0{\left(l_{ik}\right)}\right)^{1-z_{ik}},
    \end{split}\\
    \begin{split}\propto {}&\exp\Vast\{-\frac{\hat{a}_{\tau_i}}{2\hat{b}_{\tau_i}}\Eover{\mathbf{L}_{-ik},\mathbf{F},\mathbf{Z}_{-ik}}{-2\boldsymbol{y}_{i\cdot}^{\mathsf{T}}\boldsymbol{f}_{k\cdot} l_{ik} + 2\sum_{k'\neq k} z_{ik'}\boldsymbol{f}_{k}^{\mathsf{T}}\boldsymbol{f}_{k'\cdot}l_{ik'}l_{ik} + \boldsymbol{f}_{k}^{\mathsf{T}}\boldsymbol{f}_{k}l_{ik}^2}\\
    &\hspace{3em} +\frac{1}{2}\left(\psi{\left(\hat{a}_{\alpha_k}\right)}-\log 2\pi\hat{b}_{\alpha_k}-\frac{\hat{a}_{\alpha_k}}{\hat{b}_{\alpha_k}}l_{ik}^2\right) \Vast\}^{z_{ik}}\times \pi_k^{z_{ik}}\left(\left(1-\pi_k\right)\delta_0{\left(l_{ik}\right)}\right)^{1-z_{ik}}
    \end{split}\\
    \begin{split}\propto {}&\exp\Vast\{\frac{\hat{a}_{\tau_i}}{\hat{b}_{\tau_i}}\left(-\sum_{j=1}^N \left( y_{ij}\mu_{f_{kj}} - \sum_{k'\neq k} \eta_{ik'}\mu_{f_{kj}}\mu_{f_{k'j}}\mu_{l_{ik'}} \right)l_{ik} -\sum_{j=1}^N\left(\mu^2_{f_{kj}}+\sigma^2_{f_{kj}}\right) \frac{l_{ik}^2}{2}\right)\\
    &\hspace{3em} +\frac{1}{2}\left(\psi{\left(\hat{a}_{\alpha_k}\right)}-\log 2\pi\hat{b}_{\alpha_k}-\frac{\hat{a}_{\alpha_k}}{\hat{b}_{\alpha_k}}l_{ik}^2\right) \Vast\}^{z_{ik}}\times \pi_k^{z_{ik}}\left(\left(1-\pi_k\right)\delta_0{\left(l_{ik}\right)}\right)^{1-z_{ik}},
    \end{split}
\end{align*}
which corresponds to the updates
\begin{align*}
\begin{split}
    \sigma^{2*}_{l_{ik}} &= \left(\frac{\hat{a}_{\tau_i}}{\hat{b}_{\tau_i}}\sum_{j=1}^N\left(\mu^2_{f_{kj}}+\sigma^2_{f_{kj}}\right)+\frac{\hat{a}_{\alpha_k}}{\hat{b}_{\alpha_k}}\right)^{-1}\\
    \mu_{l_{ik}}^* &= \frac{\hat{a}_{\tau_i}}{\hat{b}_{\tau_i}}\sigma^{2*}_{l_{ik}}
    \sum_{j=1}^N \left( y_{ij}\mu_{f_{kj}} - \sum_{k'\neq k} \eta_{ik'}\mu_{f_{kj}}\mu_{f_{k'j}}\mu_{l_{ik'}} \right)
\end{split}\\
q{\left(z_{ik}\right)}
&\propto 
\exp{\Biggl\{\frac{z_{ik}}{2}\left(\psi{\left(\hat{a}_{\alpha_k}\right)}-\log 2\pi\hat{b}_{\alpha_k} + \frac{\mu_{l_{ik}}^{2*}}{\sigma^{2*}_{l_{ik}}}\right) \Biggr\}}
\left(\sqrt{2\pi\sigma^{2*}_{l_{ik}}} \pi_k\right)^{z_{ik}}\left(1-\pi_k\right)^{1-z_{ik}}.
\end{align*}
Coordinate ascent on $f_{kj}$ gives
\begin{align*}
q^*{\left(f_{kj}\right)}
    \propto {}& \exp{\left\{\Eover{\mathbf{L},\mathbf{F}_{-kj},\mathbf{Z},\boldsymbol{\tau},\boldsymbol\alpha}{\log p\giventhat*{f_{kj}}{\mathbf{Y},\mathbf{L},\mathbf{F}_{-kj},\mathbf{Z},\boldsymbol{\tau},\boldsymbol\alpha}}\right\}}\\
    \propto {}& \exp{\left\{\Eover{\mathbf{L},\mathbf{F}_{-kj},\mathbf{Z},\boldsymbol{\tau}}{-\frac{1}{2}\left(\boldsymbol{y}_{\cdot j}-\mathbf{L}\boldsymbol{f}_{\cdot j}\right)^{\mathsf{T}}D_{\boldsymbol{\tau}}\left(\boldsymbol{y}_{\cdot j}-\mathbf{L}\boldsymbol{f}_{\cdot j}\right)}-\frac{1}{2}f_{kj}^2\right\}}\\
    \propto {}&
    \exp{\left\{\boldsymbol{y}_{\cdot j}^{\mathsf{T}}D_{\mean{\boldsymbol{\tau}}}\mean{\boldsymbol{l}_{\cdot k}}f_{kj}
    -\sum_{k'\neq k} \mean{f_{k'j}\boldsymbol{l}_{\cdot k'}^\mathsf{T}D_{\boldsymbol{\tau}}\boldsymbol{l}_{\cdot k}}f_{kj}
    -\frac{1}{2}\left(\mean{\boldsymbol{l}_{\cdot k}^\mathsf{T}D_{\boldsymbol{\tau}}\boldsymbol{l}_{\cdot k}}+1\right)f_{kj}^2\right\}}
\end{align*}
where
\begin{align*}
    D_{\mean{\boldsymbol\tau}} &=\text{diag}{\left(\left\{\frac{\hat{a}_{\tau_i}}{\hat{b}_{\tau_i}}\right\}_{i=1}^G\right)}\\
    \mean{\boldsymbol{l}_{\cdot k}} &=\left\{\eta_{ik}\mu_{l_{ik}}\right\}_{i=1}^G\\
    \mean{f_{k'j}\boldsymbol{l}_{\cdot k'}^\mathsf{T}D_{\boldsymbol{\tau}}\boldsymbol{l}_{\cdot k}}
    &= \mu_{f_{k'j}} \sum_{i=1}^G \frac{\hat{a}_{\tau_i}}{\hat{b}_{\tau_i}}\eta_{ik}\eta_{ik'}\mu_{l_{ik}}\mu_{l_{ik'}}\\
    \mean{\boldsymbol{l}_{\cdot k}^\mathsf{T}D_{\boldsymbol{\tau}}\boldsymbol{l}_{\cdot k}}
    &= \sum_{i=1}^G \frac{\hat{a}_{\tau_i}}{\hat{b}_{\tau_i}}\eta_{ik}\left(\mu_{l_{ik}}^2+\sigma_{l_{ik}}^2\right),
\end{align*}
which corresponds to the updates
\begin{align*}
\begin{split}
    \sigma_{f_{kj}}^{2*} &= \left(\mean{\boldsymbol{l}_{\cdot k}^\mathsf{T}D_{\boldsymbol{\tau}}\boldsymbol{l}_{\cdot k}} + 1\right)^{-1}\\
    \mu_{f_{kj}}^* &= \sigma_{f_{kj}}^{2*}\left(\boldsymbol{y}_{\cdot j}^{\mathsf{T}}D_{\mean{\boldsymbol{\tau}}}\mean{\boldsymbol{l}_{\cdot k}}
    -\sum_{k'\neq k} \mean{f_{k'j}\boldsymbol{l}_{\cdot k'}^\mathsf{T}D_{\boldsymbol{\tau}}\boldsymbol{l}_{\cdot k}}\right).
\end{split}
\end{align*}
Coordinate ascent on $\tau_i$ gives
\begin{align*}
    q^*{\left(\tau_i\right)}
    &\propto
    \exp{\left\{\Eover{\mathbf{L},\mathbf{F},\mathbf{Z},\boldsymbol\alpha}{\log p\giventhat*{\tau_i}{\mathbf{Y},\mathbf{L},\mathbf{F},\mathbf{Z},\boldsymbol\alpha}}\right\}}\\
    &\propto
    \exp{\left\{\left(a_{\tau}-1+\frac{N}{2}\right)\log \tau_i-b_{\tau}\tau_i - \frac{\tau_i}{2}\Eover{\mathbf{L},\mathbf{F},\mathbf{Z}}{\left(\boldsymbol{y}_{i\cdot}-\mathbf{F}^{\mathsf{T}}\boldsymbol{l}_{i\cdot}\right)^{\mathsf{T}}\left(\boldsymbol{y}_{i\cdot}-\mathbf{F}^{\mathsf{T}}\boldsymbol{l}_{i\cdot}\right)}\right\}}\\
    &\propto
    \exp{\left\{\left(a_{\tau}-1+\frac{N}{2}\right)\log \tau_i-\left(b_{\tau} + \frac{1}{2}\left(\boldsymbol{y}_{i\cdot}^{\mathsf{T}}\boldsymbol{y}_{i\cdot}-2\mean{\boldsymbol{l}_{i\cdot}}^{\mathsf{T}}\mean{\mathbf{F}}\boldsymbol{y}_{i\cdot}+\mean{\boldsymbol{l}_{i\cdot}^{\mathsf{T}}\mathbf{F}\mathbf{F}^{\mathsf{T}}\boldsymbol{l}_{i\cdot}}\right)\right)\tau_i\right\}}
\end{align*}
where
\begin{align*}
    \mean{\boldsymbol{l}_{i\cdot}} &=\left\{\eta_{ik}\mu_{l_{ik}}\right\}_{k=1}^K\\
    \Bigl[\mean{\mathbf{F}}\Bigr]_{kj}
    &= \mu_{f_{kj}}\\
    \mean{\boldsymbol{l}_{i\cdot}^{\mathsf{T}}\mathbf{F}\mathbf{F}^{\mathsf{T}}\boldsymbol{l}_{i\cdot}}
    &= \sum_{k=1}^K\sum_{k'=1}^K\left(\eta_{ik}\eta_{ik'}^{1-\delta_{kk'}}\left(\mu_{l_{ik}}\mu_{l_{ik'}}+\delta_{kk'}\sigma_{l_{ik}}^2\right)\sum_{j=1}^N\left(\mu_{f_{kj}}\mu_{f_{k'j}}+\delta_{kk'}\sigma^2_{f_{kj}}\right)\right),
\end{align*}
which corresponds to the updates
\begin{align*}
\begin{split}
    \hat{a}_{\tau_i}^* &= a_{\tau} + \frac{N}{2}\\
    \hat{b}_{\tau_i}^* &= b_{\tau} + \frac{1}{2}\left(\boldsymbol{y}_{i\cdot}^{\mathsf{T}}\boldsymbol{y}_{i\cdot}-2\mean{\boldsymbol{l}_{i\cdot}}^{\mathsf{T}}\mean{\mathbf{F}}\boldsymbol{y}_{i\cdot}+\mean{\boldsymbol{l}_{i\cdot}^{\mathsf{T}}\mathbf{F}\mathbf{F}^{\mathsf{T}}\boldsymbol{l}_{i\cdot}}\right).
\end{split}
\end{align*}
Coordinate ascent on $\alpha_k$ gives
\begin{align*}
    q^*{\left(\alpha_k\right)}
    &\propto
    \exp{\left\{\Eover{\mathbf{L},\mathbf{F},\mathbf{Z},\boldsymbol\tau}{\log p\giventhat*{\alpha_k}{\mathbf{Y},\mathbf{L},\mathbf{F},\mathbf{Z},\boldsymbol\tau}}\right\}}\\
    &\propto
    \exp{\left\{\left(a_{\alpha}-1+\frac{1}{2}\Eover{\mathbf{Z}}{\sum_{i=1}^G z_{ik}}\right)\log \alpha_k - b_{\alpha}\alpha_k - \frac{\alpha_k}{2}\Eover{\mathbf{L},\mathbf{Z}}{\sum_{i\cl z_{ik} = 1} l_{ik}^2}\right\}}
\end{align*}
which corresponds to the updates
\begin{align*}
\begin{split}
    \hat{a}_{\alpha_k}^* &= a_{\alpha} + \frac{1}{2}\sum_{i=1}^G \eta_{ik}\\
    \hat{b}_{\alpha_k}^* &= b_{\alpha} + \frac{1}{2}\sum_{i=1}^G \eta_{ik}\left(\sigma^2_{l_{ik}}+\mu^2_{l_{ik}}\right).
\end{split}
\end{align*}

\textbf{Computing the \textsc{elbo}.} For the sparse factor model, the \textsc{elbo} is given by
$$\textsc{\textsc{elbo}}{(q)} = \Eover{\mathbf{L},\mathbf{F},\mathbf{Z},\boldsymbol{\tau},\boldsymbol\alpha}{\log p{\left(\mathbf{Y},\mathbf{L},\mathbf{F},\mathbf{Z},\boldsymbol{\tau},\boldsymbol\alpha\right)} - \log q{\left(\mathbf{L},\mathbf{F},\mathbf{Z},\boldsymbol{\tau},\boldsymbol\alpha\right)}}.$$
Breaking this down into components, the expectation of the first logarithm consists of
\begin{align*}
    \Eover{\mathbf{L},\mathbf{F},\boldsymbol{\tau}}{\log p\giventhat{y_{ij}}{\mathbf{L},\mathbf{F},\boldsymbol{\tau}}}
    ={}& \frac{1}{2}\Eover{\mathbf{L},\mathbf{F},\boldsymbol{\tau}}{\log \frac{\tau_i}{2\pi} - \tau_i\left(y_{ij}-\boldsymbol{l}_{i\cdot}^\mathsf{T}\boldsymbol{f}_{\cdot j}\right)^2}\\
    ={}& \frac{1}{2}\Biggl( \psi{\left(\hat{a}_{\tau_i}\right)}-\log 2\pi\hat{b}_{\tau_i} -\frac{\hat{a}_{\tau_i}}{\hat{b}_{\tau_i}}\biggl( -2y_{ij}\sum_{k=1}^K\eta_{ik}\mu_{l_{ik}}\mu_{f_{kj}}\\
    &\hspace{12em} +y_{ij}^2    +\mean{\left(\boldsymbol{l}_{i\cdot}^\mathsf{T}\boldsymbol{f}_{\cdot j}\right)^2} \biggr) \Biggr)\\
    \Eover{\mathbf{L},\mathbf{Z},\boldsymbol{\alpha}}{\log p\giventhat{l_{ik}}{\mathbf{Z},\boldsymbol{\alpha}}}
    ={}& \Eover{\mathbf{L},\mathbf{Z},\boldsymbol{\alpha}}{\frac{z_{ik}}{2}\left(\log\frac{\alpha_k}{2\pi} -\alpha_k l_{ik}^2 \right) + \left(1 - z_{ik}\right) \log \delta_0{(l_{ik})}}\\
    ={}& \frac{\eta_{ik}}{2}\left(\psi{\left(\hat{a}_{\alpha_k}\right)}-\log 2\pi\hat{b}_{\alpha_k} -\frac{\hat{a}_{\alpha_k}}{\hat{b}_{\alpha_k}}\left(\mu_{l_{ik}}^2+\sigma_{l_{ik}}^2\right)\right)\\
    &\hspace{1em} +\Eover{\mathbf{L},\mathbf{Z}}{\left(1 - z_{ik}\right) \log \delta_0{(l_{ik})}}\\
    \Eover{\mathbf{Z}}{\log p{(z_{ik})}}
    ={}& \eta_{ik}\log\pi_k + \left(1-\eta_{ik}\right)\log\left(1-\pi_k\right)\\
    \Eover{\mathbf{F}}{\log p{(f_{kj})}}
    ={}& -\frac{1}{2}\left(\mu_{f_{kj}}^2+\sigma_{f_{kj}}^2 + \log 2\pi\right)\\
    \Eover{\boldsymbol{\tau}}{\log p{(\tau_i)}}
    ={}& \Eover{\boldsymbol{\tau}}{\left(a_\tau-1\right)\log \tau_i - b_\tau\tau_i} + a_\tau\log b_\tau - \log\Gamma{(a_\tau)}\\
    ={}& \left(a_\tau-1\right)\left(\psi{\left(\hat{a}_{\tau_i}\right)}-\log \hat{b}_{\tau_i}\right) - \frac{\hat{a}_{\tau_i}}{\hat{b}_{\tau_i}}b_\tau +  a_\tau\log b_\tau - \log\Gamma{(a_\tau)}\\
    \Eover{\boldsymbol{\alpha}}{\log p{(\alpha_k)}}
    ={}& \Eover{\boldsymbol{\alpha}}{\left(a_\alpha-1\right)\log \alpha_k - b_\alpha\alpha_k} +  a_\alpha\log b_\alpha - \log\Gamma{(a_\alpha)}\\
    ={}& \left(a_\alpha-1\right)\left(\psi{\left(\hat{a}_{\alpha_k}\right)}-\log \hat{b}_{\alpha_k}\right) - \frac{\hat{a}_{\alpha_k}}{\hat{b}_{\alpha_k}}b_\alpha +  a_\alpha\log b_\alpha - \log\Gamma{(a_\alpha)}
\end{align*}
where
$$
\mean{\left(\boldsymbol{l}_{i\cdot}^\mathsf{T}\boldsymbol{f}_{\cdot j}\right)^2}
= \sum_{k=1}^K\sum_{k'=1}^K\eta_{ik}\eta_{ik'}^{1-\delta_{kk'}}\left(\mu_{l_{ik}}\mu_{l_{ik'}}+\delta_{kk'}\sigma_{l_{ik}}^2\right)\left(\mu_{f_{kj}}\mu_{f_{k'j}}+\delta_{kk'}\sigma^2_{f_{kj}}\right).
$$
Using standard differential entropy results, the expectation of the second logarithm consists of
\begin{align*}
    \Eover{\mathbf{L},\mathbf{Z}}{-\log q{(l_{ik},z_{ik})}}
    ={}& \frac{\eta_{ik}}{2}\left(\log 2\pi\sigma_{l_{ik}}^2 + 1\right) -\eta_{ik}\log\eta_{ik} -\left(1-\eta_{ik}\right)\log\left(1-\eta_{ik}\right)\\
    &\hspace{1em} -\Eover{\mathbf{L},\mathbf{Z}}{\left(1 - z_{ik}\right) \log \delta_0{(l_{ik})}}\\
    \Eover{\mathbf{F}}{-\log q{(f_{kj})}}
    ={}& \frac{1}{2}\left(\log 2\pi\sigma^2_{f_{kj}} + 1\right)\\
    \Eover{\boldsymbol{\tau}}{-\log q{(\tau_i)}}
    ={}& \hat{a}_{\tau_i} -\log\hat{b}_{\tau_i} +\log\Gamma{(\hat{a}_{\tau_i})} +\left(1 - \hat{a}_{\tau_i}\right)\psi{\left(\hat{a}_{\tau_i}\right)}\\
    \Eover{\boldsymbol{\alpha}}{-\log q{(\alpha_k)}}
    ={}& \hat{a}_{\alpha_k} -\log\hat{b}_{\alpha_k} +\log\Gamma{(\hat{a}_{\alpha_k})} +\left(1 - \hat{a}_{\alpha_k}\right)\psi{\left(\hat{a}_{\alpha_k}\right)}.
\end{align*}
The \textsc{elbo} may be calculated by summing up these expectations appropriately.

\section{Relabelling samples}\label{rs}

A relabelling algorithm, similar to that of \cite{invariance}, is used to deal with these model non-identifiability issues of label symmetry and sign ambiguity. For the sparse factor model, a relabelling consists of permuting the factors, and potentially flipping the signs of all entries of some factors. Following the method of \cite{relabel}, a decision-theoretic approach is to define a loss function for a set of actions and relabellings, and select the action and relabelling which minimises the posterior expected loss. This is done with the aim of relabelling samples such that they correspond to being sampled around the same mode. Define an action
\begin{equation*}
    \mathbf{a} = \left(
    \left\{m_{f_{kj}}\right\}_{\substack{j=1:N\\k=1:K}},
    \left\{s^2_{f_{kj}}\right\}_{\substack{j=1:N\\k=1:K}}
    \right)
\end{equation*} to be a choice of means and variances of the entries of $\mathbf{F}$. Let $\sigma\in S_K$ and $\boldsymbol{\nu}\in \{-1,1\}^K$, where $S_K$ is the set of permutations on the set $\{1,2,\ldots,K\}$. Define a loss function as follows:
\begin{equation*}
    \mathcal{L}{(\mathbf{a},\sigma,\boldsymbol{\nu};\mathbf{F})}
    = -\sum_{k=1}^K \sum_{j=1}^N \mathcal{N}\giventhat{\nu_{\sigma{(k)}} f_{\sigma(k)j}}{m_{f_{kj}},s^2_{f_{kj}}}.
\end{equation*}

Suppose $T$ simulated samples of $\mathbf{F}$, namely $\left\{\mathbf{F}^{(t)}\right\}_{t=1:T}$ are to be relabelled, which may be obtained from multiple chains. The actions and relabellings $\mathbf{a}$ and $\left\{\left(\sigma^{(t)},\boldsymbol{\nu}^{(t)}\right)\right\}_{t=1:T}$ are to be chosen such that the Monte Carlo risk
\begin{equation*}
    \mathcal{R}_{\text{MC}} = \sum_{t=1}^T \mathcal{L}{\left(\mathbf{a},\left\{\left(\sigma^{(t)},\boldsymbol{\nu}^{(t)}\right)\right\}_{t=1:T};\mathbf{F}^{(t)}\right)}
\end{equation*}
is minimised. In the case of multiple chains, it may be more appropriate to first scale each row of $\mathbf{F}$ to unit norm. After initialising $\mathbf{a}$ and $\left\{\left(\sigma^{(t)},\boldsymbol{\nu}^{(t)}\right)\right\}_{t=1:T}$, a local optimum may be obtained by alternating between the following steps:

\begin{addmargin}[2em]{2em}
\begin{enumerate}
    \item Given the current values of $\left\{\left(\sigma^{(t)},\boldsymbol{\nu}^{(t)}\right)\right\}_{t=1:T}$, choose $\mathbf{a}$ such that the Monte Carlo risk is minimised.
    \item Given the current action $\mathbf{a}$, choose $\left\{\left(\sigma^{(t)},\boldsymbol{\nu}^{(t)}\right)\right\}_{t=1:T}$ such that the Monte Carlo risk is minimised.
\end{enumerate}
\end{addmargin}
This procedure is terminated when a fixed point is reached. The final signflips and permutations $\left\{\left(\sigma^{(t)},\boldsymbol{\nu}^{(t)}\right)\right\}_{t=1:T}$ are then applied to all relevant variables simulated. 

Step 1 may be solved analytically, by setting partial derivatives of the Monte Carlo risk with respect to the action parameters to zero. This is equivalent to finding the maximum likelihood estimators, summarised by the following updates:
\begin{align*}
    \widehat{m_{f_{kj}}} &= \frac{1}{T}\sum_{t=1}^T \nu^{(t)}_{\sigma^{(t)}{(k)}} f_{\sigma^{(t)} (k)j}^{(t)}\\
    \widehat{s^2_{f_{kj}}} &= \frac{1}{T}\sum_{t=1}^T \left(\nu^{(t)}_{\sigma^{(t)}{(k)}} f_{\sigma^{(t)} (k)j}^{(t)} - \widehat{m_{f_{kj}}}\right)^2. 
\end{align*}

Step 2 is equivalent to the linear assignment problem. For each simulated sample, this may be solved by an $\mathcal{O}{(K^3)}$ algorithm of \cite{lap} after a cost matrix is constructed. The construction of the cost matrix itself takes $\mathcal{O}{\left(K^2(G+N)\right)}$ time (for each simulated sample).







\end{appendices}

\bibliographystyle{apalike}
\bibliography{reference}

\end{document}